\documentstyle[12pt, aaspp4]{article}
\lefthead{FRYER \& WOOSLEY}
\righthead{}
\begin{document}
\begin{center}
Accepted by {\em The Astrophysical Journal}
\end{center}
\vspace{1.cm}
\title{Gamma-Ray Bursts From Neutron Star Phase Transitions}
\author{Chris L. Fryer \& S. E. Woosley}
\affil{Lick Observatory, University of California Observatories, \\ 
Santa Cruz, CA 95064 }
\affil{Max Planck Institut f\"ur Astrophysik, Garching, 85740,
Germany \\ cfryer@ucolick.org, woosley@ucolick.org}
\authoremail{cfryer@ucolick.org;\  woosley@ucolick.org} 

\begin{abstract} 
The phase-transition induced collapse of a neutron star to a more compact
configuration (typically a ``strange'' star) and the subsequent core
bounce is often invoked as a model for gamma-ray bursts.  We present
the results of numerical simulations of this kind of event using
realistic neutrino physics and a high density equation of
state. The nature of the collapse itself is represented by the
arbitrary motion of a piston deep within the star, but if
any shock is to develop, the transition, or at least its final stages,
must occur in less than a sonic time. Fine surface zoning is employed
to adequately represent the acceleration of the shock to relativistic
speeds and to determine the amount and energy of the ejecta.  We find
that these explosions are far too baryon-rich ($M_{\rm ejecta} \sim
0.01 M_{\odot}$) and have much too low an energy to explain gamma-ray
bursts. The total energy of the ejecta having relativistic $\Gamma
\gtrsim 40$ is less than $10^{46}$erg even in our most optimistic
models (deep bounce, no neutrino losses or photodisintegration).
However, the total energy of all the ejecta, mostly mildly
relativistic, is $\sim 10^{51}$erg and, if they occur, these events
might be observed.  They would also contribute to Galactic
nucleosynthesis, especially the $r$-process, even though the most
energetic layers are composed of helium and nucleons, not heavy
elements.
\end{abstract}


\section{Introduction}

A major goal in modern model building for cosmological gamma-ray
bursts is finding a source which provides both the high energies
($\gtrsim 10^{51}$erg for symmetric explosions) and high Lorentz
factors ($\Gamma \gtrsim 100$) required to explain the observations
(e.g., M\'esz\'aros \& Rees 1993).  One often proposed source for this
energy is an explosion resulting from the phase transition of a neutron
star to a ``strange'' or ``hybrid'' star (Ramaty et al. 1980; Ramaty,
Lingenfelter \& Bussard 1981; Brecher 1982; Ellison \& Kazanas 1983;
Bonazzola 1986; Michel 1988; Haensel, Paczynski \& Amsterdaamski 1991;
Ma \& Xie 1996; Ma \& Luo 1996; Shaviv \& Dar 1996, Qin et al. 1997).
Though there is some variation in the models - sometimes a critical mass
is achieved for an accreting neutron star, sometimes only a single star
is involved - all take note of the large gravitational binding energy
of a neutron star and speculate that some fraction of this can be
tapped and converted into an outgoing shock wave when the inner core
abruptly makes the transition to a more compact state. Because the
density is so high, neutrino losses are small, except very near the
surface; similarly photodisintegration losses are negligible, so the
``prompt shock'' mechanism that fails to give mass ejection in
standard supernova models (e.g. Bethe 1990), might, in this case, deliver
large amounts of momentum to the surface layers.  Models for strange
or ``hybrid'' stars predict central core densities nearly 5 times
greater and radii 10-20\% smaller than neutron stars of the same mass
(Rosenhauer, Staubo, \& Csernai 1991).
A rough estimate of the potential energy released, $E \sim \frac{G
M^2}{R} \frac{\Delta R}{R}$, predicts energies on the order of
$10^{52}$erg, easily sufficient for gamma-ray burst models (e.g., Ma \& Xie
1996).

However, to get the required high Lorentz factors ($\Gamma$), this
energy must be concentrated into a thin layer near the neutron star
surface so that only a small amount of mass is ejected.\footnote{To
reach a mean $\Gamma$ of 100, a $2 \times 10^{51}$erg explosion must
eject less than $10^{-5} M_{\odot}$.}  Competing with this sink for
the gravitational energy is an increase in the internal energy
that occurs when matter moves to higher gravitational potential, so
the shock does not carry an energy equal to the entire change in
gravitational potential. And of course the matter ejected will have a
distribution of kinetic energies with most of the mass concentrated at
low energies. In this paper, we show that while the phase-transitions of
neutron stars can indeed impart total energies to their ejecta in
excess of $10^{50}$erg, they fail to deposit enough energy at high
$\Gamma$s to explain gamma-ray bursts.  The phase-transition induced
collapse of neutron stars is therefore {\it not} a viable gamma-ray
burst mechanism.

\section{Simulations}
To simulate the collapse and subsequent explosion of neutron stars, we
use two separate hydrodynamic codes, one for the bounce and shock
formation in the inner core and another to study shock propagation in
the outer 0.001 to 0.01$M_{\odot}$ where very fine zoning must be
employed.  For the collapse and bounce phase of the simulation, we use
the relativistic supernova code described in Herant et al. (1994) and
Fryer et al. (1998) which includes the equation of state for high
density matter developed by Lattimer \& Swesty (1991), neutrino
processes and transport, general-relativistic effects and nuclear 
burning assuming nuclear statistical equilibrium in a Lagrangian
hydrodynamics code.  We base our calculations on a $1.4 M_{\odot}$
neutron star initial model provided by Keil(1997).  The inner 50\% of
the mass is removed from the neutron star and replaced by an inner
boundary.  After the star has recovered from this remapping (the
stable radius using our code is within 5\% of the initial radius of
Keil's model), we move the inner boundary quickly, but not instantly
to a smaller radius to artificially simulate the phase
transition.\footnote{We vary both the depth of the final inner radius
and the speed at which it drops.}  The outer zones in this calculation
are resolved down to $10^{-4} M_{\odot}$ at the edge of the neutron
star using over 450 variably massed zones.  The ultimate results of 
this calculation are an accurate estimate of the total mass ejected 
and a history of the inner boundary of this ejecta ($r(t)$), 
which we can use as a piston for the high resolution models of 
the shock progression through the neutron star crust.  

To model the shock progression through this outer crust, we use the
hydrodynamical code KEPLER (Weaver, Zimmerman, and Woosley
1978). Densities are sufficiently low ($\lesssim 10^{12}$ g cm$^{-3}$)
in this material that an ordinary equation of state (ions, radiation,
electrons and pairs of arbitrary relativicity and degeneracy) can be
used. Although the code is Newtonian, not relativistic, it has the
advantage that very fine mass zoning can be employed while conserving
momentum and energy to high accuracy. The code also has the ability to
compute approximate nucleosynthesis using a network of 19 isotopes.  

>From our bounce simulations, we derive a piston velocity for the outer
$0.01-0.001 M_{\odot}$ of ejecta (depending upon the model) which is
then placed at the inner boundary of the KEPLER simulation.  Using
$\sim 200$ logarithmically binned zones, we are able to compute shock
velocities and nucleosynthetic yields with an outer resolution of
$10^{-11} M_{\odot}$.  Later in the paper we shall discuss the 
validity of our Newtonian results for a problem where the motion 
is trans-relativistic.

\subsection{Bounce Results}
Table 1 summarizes the results of a series of simulations in which we
varied the depth and speed at which the inner boundary was lowered.  
We assume that the inner boundary in these simulations moves in 
rapidly (with its free-fall velocity) and have varied the strength of the 
bounce.  If the material collapses much more slowly, the bounce 
will be much weaker and explosions may not occur at all.  

A common misconception in gamma-ray burst work is that all of the
potential energy gained during the compression of a compact object is
immediately injected into the exploding material.  From the mass-point
trajectories of our most energetic explosion (Figure 1)\footnote{In
this model, the inner boundary falls inward from 7.5km to 4.5km.  The
inner boundary is then given a kinetic energy per gram 
equivalent to the potential energy released and it
drives the shock through the infalling zones.} we see that, although
there may be a great deal of potential energy released as the neutron
star converts to a strange star, only a small fraction of this
potential energy appears in the ejecta.  The kinetic energy of the
inner boundary is almost entirely reprocessed into potential energy as
the outer 40\% of the neutron star expands (see Figure 1).  This
energy slowly leaks out of the neutron star as it re-experiences
Kelvin-Helmholtz evolution.  This effect is akin to the process in
core-collapse supernovae where, during the explosion, the
proto-neutron star collapses down to 50km, driving an explosion with
only 20\% of the gravitational energy that it will ultimately lose in
becoming a cold 10km neutron star.  Hence, simple estimates of the
energy, and nucleosynthetic yield of these explosions.  such as those
by Ma \& Xie (1996), begin by overestimating the available energy by a
factor of 5-10.

A more crucial misconception is the assumption that the explosion
ejects only a tiny fraction of the collapsing object.  In table 1, we
see that the mass ejected ranges from $0.001-0.2 M_{\odot}$
(corresponding to 10-100 zones in our bounce simulations), much larger than
gamma-ray burst models require.  Although we can increase the
explosion energy by altering the depth at which we lower the inner
boundary, we do not change the shock velocity
significantly, but rather eject more material.  Note that if the
neutron star collapses to a black hole, no matter is ejected, the
neutrinos are trapped in the collapse and no energy is ejected.  Thus,
the collapse model proposed by Qin et al. (1997) fails to produce any
explosion, let alone a gamma-ray burst.

Neutrinos play a minor role in the mass ejection 
in these objects simply because, by the time the 
ejecta begins to get optically thin, 90\% of the 
energy has already been converted to kinetic energy.
The opacity for neutrinos due to absorption onto free 
nucleons is (Herant et al. 1994):
\begin{equation}
\kappa=8.5 \times 10^{-18} \left( \frac{\epsilon_{\nu}}
{12.5 \rm{MeV}} \right)^{2} \; {\rm cm^2/g},
\end{equation}
where $\epsilon_{\nu}$ is the neutrino energy.  
The optical depth ($\tau=\kappa \rho \Delta R$) 
becomes one when the shock has reached 50 km and the 
density is $\sim 10^{10}$g/cm$^3$.  

After the initial ejection of matter, neutrinos 
will continue to deposit energy outside the neutrinosphere 
due to neutrino/anti-neutrino pair annihilation.  
We reiterate the fact that neutrino/anti-neutrino 
annihilation requires nearly head-on neutrino 
collisions (Janka 1991) and will not be effective 
in driving a gamma-ray burst for our spherically 
symmetric explosions\footnote{Angular momentum 
does not alter our results.  The outer crust of 
the neutron star does not collapse significantly 
before being ejected and their trajectories 
are not significantly effected by angular momentum.  
Likewise, the explosion is too rapid to allow the 
development of convection and it will also 
not effect the results.}.  
The energy deposition near the neutrinosphere drives 
a wind which is far too baryon rich to be a gamma-ray 
burst model (Duncan et al. 1986).  This effect has 
been seen in simulations for neutron star mergers 
(Janka \& Ruffert 1996) and the accretion induced 
collapse of white dwarfs (Woosley \& Baron 1992, 
Fryer et al. 1998) and the resultant winds are non-relativistic.  
However, these winds are likely to eject an 
additional $0.01 M_{\odot}$ of possibly neutron-rich 
material.

\subsection{Shock Progression Through The Crust}
As the shock progresses through the outer crust of the neutron star,
it accelerates.  For non-relativistic shocks, the shock velocity is
given by (Sedov 1959):
\begin{equation}
v_{shock} \propto t^{\frac{\omega-3}{5-\omega}},
\end{equation}
where $\omega$ is given by the density structure of the medium through
which the shock travels ($\rho \propto r^{-\omega}$).  Shocks 
propogating through density profiles where $\omega>3$ accelerate.  
As the shock propogates through the exponentially decreasing density 
profile of the neutron star crust, it accelerates and quickly becomes 
relativistic.  Analytic
solutions of ultra-relativistic shocks and semi-analytic solutions of
mildly relativistic shocks also exist (Johnson \& McKee 1971; McKee \&
Colgate 1973; Gnatyk 1985).  For an exponentially falling density
profile (to high accuracy, the structure of the outer crust of a
neutron star is an exponential atmosphere) the product of the Lorentz
factor ($\Gamma$) and the velocity of the shock ($\beta=v/c$ where $c$
is the speed of light) is given by (Gnatyk 1985):
\begin{equation}
\Gamma \beta \propto (\rho r^{N+1})^{-\alpha},
\end{equation}
where $N$ is a geometric factor set to 2 for spherical symmetry, and
$\alpha$ is determined, via simulations, to be $\sim$0.20.

Unfortunately, this solution does not apply to the velocity of the
material after the shock has passed through the material and the
internal energy of the material begins to convert back to kinetic
energy\footnote{The final velocity depends upon the equation of state
and the effects of gravity, effects not modeled by the semi-analytic
solution}.  Therefore, we are forced to simulate, to high spatial
resolution, the velocity of the ejecta.  For this purpose, we use the
code KEPLER.  Although KEPLER is not a relativistic code, we can
determine the effective Lorentz factor of the expanding material by setting
$\Gamma=E_{tot}/(M c^2)$ where $E_{tot}$ is the total energy from the
simulation and $M$ is the mass of the ejecta.  Figure 2 plots 
the energy output of our explosions as a function of material 
ejected with Lorentz factors greater than $\Gamma_0$.  Comparing our
simulations just after shock breakout with the analytic 
solutions shows that this conversion holds quite well at 
this point in the simulation (see Figure 2).

However, the material continues to accelerate after the shock 
breaks out of the neutron star as internal energy is converted 
to kinetic energy.  Although energy is conserved globally, our 
conversion for $\Gamma$ places more energy than the relativistic 
solution at high Lorentz factors, overestimating the amount of 
highly relativistic material at late times (see appendix).  
The exact energy output of these explosions, then,  
is bounded by the energy distribution given at shock breakout and our 
overestimate of the {\sl final} energy output.  A more accurate distribution 
requires relativistic simulations.  Assuming the 
upper limit given by our simulations, there is still very little energy 
($<10^{46}erg$) ejected at$\Gamma$'s greater than 40 (see Figure
2).  Thus, even in our most energetic explosion (which is probably much 
stronger than reality), there is insufficient
energy at high Lorentz factors to explain gamma-ray bursts.

\section{Observational Properties}
The primary result of this work is simply that the energies produced via the
phase-transition of neutron stars at high Lorentz factors ($\Gamma >
40$) is at least 4-5 orders of magnitude too low to explain
cosmological gamma-ray bursts, and these explosions can not explain
the gamma-ray burst observations.
Although the explosion resulting from a phase-transition of a neutron star into
a strange star is not sufficiently relativistic to be a gamma-ray
burst source, it does have a lot of energy ($\sim 10^{51}$erg) at
mildly relativistic velocities.  To estimate the luminosity profile of
these explosions, we assume the ejecta loses its energy over its
deceleration timescale (Rees \& M\'esz\'aros 1992):
\begin{equation}
L=E(\Gamma)/t_{\rm dec}(\Gamma)
\end{equation}
where the deceleration timescale ($t_{\rm dec}$) is given by:
\begin{equation}
t_{\rm dec}(\Gamma)=\frac{r_{\rm dec}(\Gamma)}{2 \Gamma^2 c}
\end{equation}
and the deceleration radius ($r_{\rm dec}$) is: 
\begin{equation}
r_{\rm dec}(\Gamma)=\left ( \frac{E(\Gamma)}{4/3 \pi \rho_{\rm ext} 
\Gamma^2} \right )^{1/3}.
\end{equation}
In these equations, $\Gamma$ is the Lorentz factor, $E(\Gamma)$ is the
explosion energy at that Lorentz factor, and $\rho_{\rm ext}$ is the
density of the interstellar medium (for our purposes, we assume a
number density of 1 cm$^{-3}$).  Under these assumptions, we can
estimate the luminosity of phase-transition induced explosions of
neutron stars (see Figure 3).  There is a small peak at early times,
but the luminosity ``burst'' quickly levels off at luminosities as
high as $10^{45}$erg/s and persists for roughly 30 days.  
Note that this luminosity is brighter than 
a typical galaxy, but 6 orders of magnitude less than a 
supernova.  Even if we assumed that the all neutron stars 
undergo such phase transitions ($\sim 1/100$yr$^{-1}$ per 
$L_*$ galaxy), these objects would make 
only a fraction ($10^{-9}$) of any magnitude limited 
supernova survey due to their lower luminosity.  If 
instead, only those neutron stars which accrete a 
sizable amount of material ($\gtrsim 0.1 M_{\odot}$)  
undergo this phase transition, the rate falls to 
roughly the X-ray binary formation rate:   
$\lesssim 10^{-5}$yr$^{-1}$ per $L_*$ galaxy (Kalogera 
\& Webbink 1998).  We must 
rely on nucleosynthetic constraints to place limits 
on these transitions.

We follow nuclear processes throughout the explosion, assuming
the electron fraction is roughly 0.5 (Our bounce models predict that
the neutrinos will reset the electron fraction to this value) and
assuming that the triple-$\alpha$ process dominates the helium
burning.  The bulk of the ejecta from these simulations is helium (see
Figure 4).  Some iron and r-process elements are produced, but only in
the interior zones where the expansion is only mildly relativistic.
This result is an additional counter argument against those models
that require high iron fractions in the ejecta (Shaviv \& Dar 1996).

The amount of $r$-process ejecta is difficult to estimate without a
detailed calculation. In the outer layers the expansion time scale is
too short and the entropy too high to reassemble alpha particles to
iron group isotopes and above. In most of the ejected mass, as in
Figure 4 around 10$^{30}$g, elements just above the iron group will
be ejected in a strong $\alpha$-rich freeze out from nuclear
statistical equilibrium (Woosley \& Hoffman 1992). There may be rare
species here such as $^{88}$Sr, $^{89}$Y, and $^{90}$Zr that limit the
occurrences of these events (see also Woosley \& Baron 1992; Woosley et
al. 1994). For an amount of material of order 10$^{27}$ - 10$^{28}$g
(Figure 4), a classical $r$-process may occur (Hoffman, Woosley, \&
Qian 1997), though the ejecta would still be chiefly helium.  Given 
a rate for neutron star phase-transitions comparable to the X-ray 
binary formation rate ($\lesssim 10^{-5}$/yr), the small amount 
of $r$-process ejecta ($\lesssim 10^{-5} M_{\odot}$) contributes 
roughly 0.1\% of the galactic $r$-process nucleosynthesis.  Only 
if nearly every neutron star formed in the galaxy experienced 
this phase transition would the resultant ejecta dominate 
the $r$-process nucleosynthesis.  At these rates, the contribution 
to cosmic rays might also be appreciable.

\acknowledgements We are grateful to Wolfgang Keil for providing us
with a neutron star model and many useful discussions.  Also, we would
like to thank Thomas Janka and Ewald M\"uller for their helpful
insight on this problem.  We would like to thank the hospitality of
the Max-Planck Institut f\"ur Astrophysik for hosting us over the
duration of this project.  This work was supported by the NSF (AST
94-17161) and by NASA (NAG5 2843 and MIT SC A 292701). SEW also
acknowledges the support of an award from the A. V. Humboldt
Foundation that covered his research expenses while at the MPA in
Garching.

\appendix
\section{Comparison of Newtonian and Special Relativistic Equations}

In this paper we have used a Newtonian code to simulate relativistic
motion.  While we compared the results of our code with analytic 
derivations for the shock progression and found good agreement
after the shock breaks out of the neutron star, we have assumed 
without equal justification, that our Newtonian code 
remains accurate during the post-shock expansion.  Here we 
show that our results for $\Gamma \gg 1$ ejecta are an upper
limit (i.e., even less material will be ejected at
high $\Gamma$'s than in Figure 2).

After the shock has traversed the neutron star, 
matter continues to accelerate as internal energy is converted 
to kinetic energy.  Assuming no energy is lost via neutrino radiation, the 
energy conservation equation can be written as the relativistically 
correct expression (e.g. Marti and M\"uller 1994):
\begin{equation}
\frac{\partial(\rho h \Gamma^2-P-
\rho \Gamma c^2)}{\partial t}
+\bigtriangledown(\rho h \Gamma^2 v -\rho c^2
\Gamma v) = 0
\end{equation}
where $\rho$ is the density, $P$, the pressure, $v$, the velocity, 
$c$, the speed of light, $t$, the time, $\Gamma \equiv 
1/\sqrt{1-(v/c)^2}$, $h \equiv c^2 + \epsilon + P/\rho$ is the enthalpy, 
and $\epsilon$, the internal energy per unit mass. By using 
the special relativistic formula for mass conservation:
\begin{equation}
\frac{\partial(\rho \Gamma)}{\partial t} + 
\bigtriangledown(\rho \Gamma v) = 0, 
\end{equation}
we derive a reduced form of the energy conservation equation:
\begin{equation}\label{eq:sr}
\frac{\partial (\rho \epsilon)}{\partial t} + 
(1-\Gamma^{-2})\frac{\partial P}{\partial t} +
\bigtriangledown(\rho \epsilon v) + 
\bigtriangledown(P v) + 
(2 \Gamma (\rho h) - \rho c^2)\Gamma^{-2} \frac{d\Gamma}{dt}=0.
\end{equation}
The first and third term in this equation correspond 
to changes in internal energy, the second and fourth 
terms correspond to changes in pressure, and the final 
term is associated with changes in kinetic energy.
This equation can be compared to its Newtonian 
counterpart:
\begin{equation}
\frac{\partial (1/2 \rho v^2 + \rho \epsilon)}{\partial t}+
\bigtriangledown [v(1/2 \rho v^2 + \rho \epsilon + P)] = 0, 
\end{equation}
which, with a similar reduction using mass conservation 
becomes:
\begin{equation}\label{eq:newt}
\frac{\partial (\rho \epsilon)}{\partial t} + 
\bigtriangledown(\rho \epsilon v) + 
\bigtriangledown(P v) + 
\rho v \frac{dv}{dt} =0.
\end{equation}
In the limit as $\Gamma, h/c^2 \Rightarrow 1$, the
$\partial P/ \partial t$ term in \ref{eq:sr} 
disappears, and the last term in both \ref{eq:sr} and
\ref{eq:newt} become equal.
As the material expands, internal energy is converted 
to kinetic energy.  To check our estimation of $\Gamma$
($=E_{tot}/(M c^2)$) during this phase, we compare the 
dependence of the rate of change of $\Gamma$ on the rate 
of internal energy change for the case of special relativity 
(\ref{eq:sr}):
\begin{equation}\label{eq:srcp}
\left .  
\frac{d \Gamma}{dt} \right )_{\rm SR}=\frac{\Gamma 
[\partial (\rho \epsilon)/ \partial t + 
\bigtriangledown (\rho \epsilon v)]}{2 (\rho c^2 + 
\rho \epsilon + P) - \rho c^2/ \Gamma}
\end{equation}
with its Newtonian counterpart (\ref{eq:newt}):
\begin{equation}\label{eq:necpp}
\frac{\partial v}{\partial t} = \frac{[\partial (\rho \epsilon)/ 
\partial t + \bigtriangledown (\rho \epsilon v)]}{\rho v}.
\end{equation}
Using our conversion assuming all the energy has been converted 
to kinetic energy ($\Gamma=E_{tot}/(M c^2) 
\Rightarrow v^2/(2 c^2)$), we get the following transformation:  
\begin{equation}
\frac{d \Gamma}{d t_{\rm SR}} = 
\frac{c^{-2} d (v^2/2)}{\Gamma^{-1} d t_{\rm Newt}}
= \frac{\Gamma v}{c^2} \frac{d v}{d t_{\rm Newt}}
\end{equation}
and equation (\ref{eq:necpp}) becomes:  
\begin{equation}\label{eq:necp}
\left . \frac{d \Gamma}{dt} \right)_{\rm Newt}
=\frac{\Gamma [\partial (\rho \epsilon)/ 
\partial t  + \bigtriangledown (\rho \epsilon v)]}{\rho c^2}.
\end{equation}
Comparing \ref{eq:srcp} and \ref{eq:necp}, it is evident that 
although our conversion works well for low $\Gamma$ material 
where the internal energy and pressure our negligible, we will 
{\it overestimate} the increase in $\Gamma$ when $\Gamma >>1$ or
when the
internal energy and pressure become important.  This implies that
we are overestimating the amount of mass at high $\Gamma$'s,
making it even less likely that these explosions are 
gamma-ray bursts.  However, this simple comparison does 
not give an exact value for $\Gamma$.  We will 
provide piston velocities for anyone who would like to 
simulate this explosion using a relativistic code.

\begin{deluxetable}{lccc}
\tablewidth{35pc}
\tablecaption{Neutron Star Collapse Simulations}
\tablehead{ \colhead{Collapse Model} & \colhead{Ejected Mass} 
& \colhead{Kinetic Energy} & 
\colhead{Energy ($\Gamma>40$)}}

\startdata
Strong Explosion\tablenotemark{a} 
& $0.017 M_{\odot}$ & $\sim 5 \times 10^{51}$ & $10^{46}$erg \nl
Weak Explosion\tablenotemark{b} 
 & $0.001 M_{\odot}$ & $\sim 3 \times 10^{50}$ & $10^{45}$erg \nl
Black Hole\tablenotemark{c}  & $0$ & $0$ & $0$ \nl

\tablenotetext{a}{In this simulation, the inner 
boundary falls inward from 7.5km to 4.5km assuming 
free-fall conditions.  It then rebounds with a kinetic 
energy per gram equal to the potential energy ($10^{53}$erg$/0.6M_{\odot}$).  
This is an overestimate of 
the energy released as some of the energy certainly goes into 
internal energy.  The inner boundary slowly decelerates 
as it expands (from gravity) until it reaches 6km, where 
it stops and is held fixed for the duration of the 
simulation.}
\tablenotetext{b}{In this simulation, the inner 
boundary falls inward from 7.5km to 5.0km assuming 
free-fall conditions.  It is then held fixed.}
\tablenotetext{c}{In this simulation, the inner 
boundary falls inward from 7.5km to 4.5km assuming
free-fall conditions.  It is then held fixed but the 
infalling cells are accreted when they reach densities 
above $5\times 10^{14}$g/cm$^3$.}

\enddata
\end{deluxetable}

\begin{figure}
\plotfiddle{figure1.ps}{6in}{-90}{70}{70}{-280}{420}
\caption{Mass-point trajectories of our most energetic 
collapse simulation.  In this simulation, the inner 
boundary falls inward from 7.5km to 4.5km assuming 
free-fall conditions.  It then rebounds with a kinetic 
energy per gram equal to the potential energy 
($10^{53}$erg$/0.6M_{\odot}$).  This is an overestimate of 
the energy released as some of the energy certainly goes into 
internal energy.  The inner boundary slowly decelerates 
as it expands (from gravity) until it reaches 6km, where 
it stops and remains constant for the duration of the 
simulation.  This simulation is almost certainly more 
energetic than we would expect from a true phase-transition 
induce collapse.  Note that almost $0.02 M_{\odot}$ is ejected 
with nearly $10^{51}$erg, but that nearly the entire star 
beyond the inner boundary expands beyond its initial 
radius (Figure 1b).  In fact, the ejecta contains less the 10\% 
of the potential energy released.}
\end{figure}

\begin{figure}
\plotfiddle{figure2.ps}{7in}{0}{70}{70}{-200}{0}
\caption{Energy of ejected material with $\Gamma \beta$ 
greater than $\Gamma_0 \beta_0$.  Both the analytic derivation and 
our simulation are compared at a time prior to shock break-out and 
the conversion of internal energy to kinetic energy.  
The two curves lie nearly 
directly on top of each other, confirming our method to convert 
from the total energy in the simulations to the value for $\Gamma$.  
Note that even for the most energetic explosion less than 
$10^{46}$ erg is ejected at $\Gamma$'s greater than 40.  The 
strong and weak explosions correspond to the strong and weak 
explosions in Table 1.}
\end{figure}

\begin{figure}
\plotfiddle{figure3.ps}{7in}{0}{70}{70}{-200}{0}
\caption{Luminosity vs. time for our explosions.  The 
luminosity remains above $10^{45}$erg/s for up to 
30 days for our most energetic explosion.  We use the 
output from the simulations, summing the contribution of 
each zone of material to the luminosity.  The highly 
relativistic material decelerates quickly and dominates the 
luminosity at short timescales whereas the massive ejecta 
at low velocities dominates the lightcurve at late times.  
Again, the strong and weak explosions correspond to the 
strong and weak explosions in Table 1.}
\end{figure}

\begin{figure}
\plotfiddle{figure4.ps}{7in}{0}{70}{70}{-200}{0}
\caption{Nucleosynthetic yield for our most energetic 
explosion.  We have assumed here that the initial electron 
fraction is roughly 0.5 (our bounce simulations predict that 
the absorbed neutrino flux resets the electron fraction to 
this value) and that the triple-$\alpha$ process is the 
dominant helium burning mechanism.  Most of the ejecta, 
especially the highly relativistic ejecta, is helium, 
but some r-process nucleosynthesis can occur.}
\end{figure}

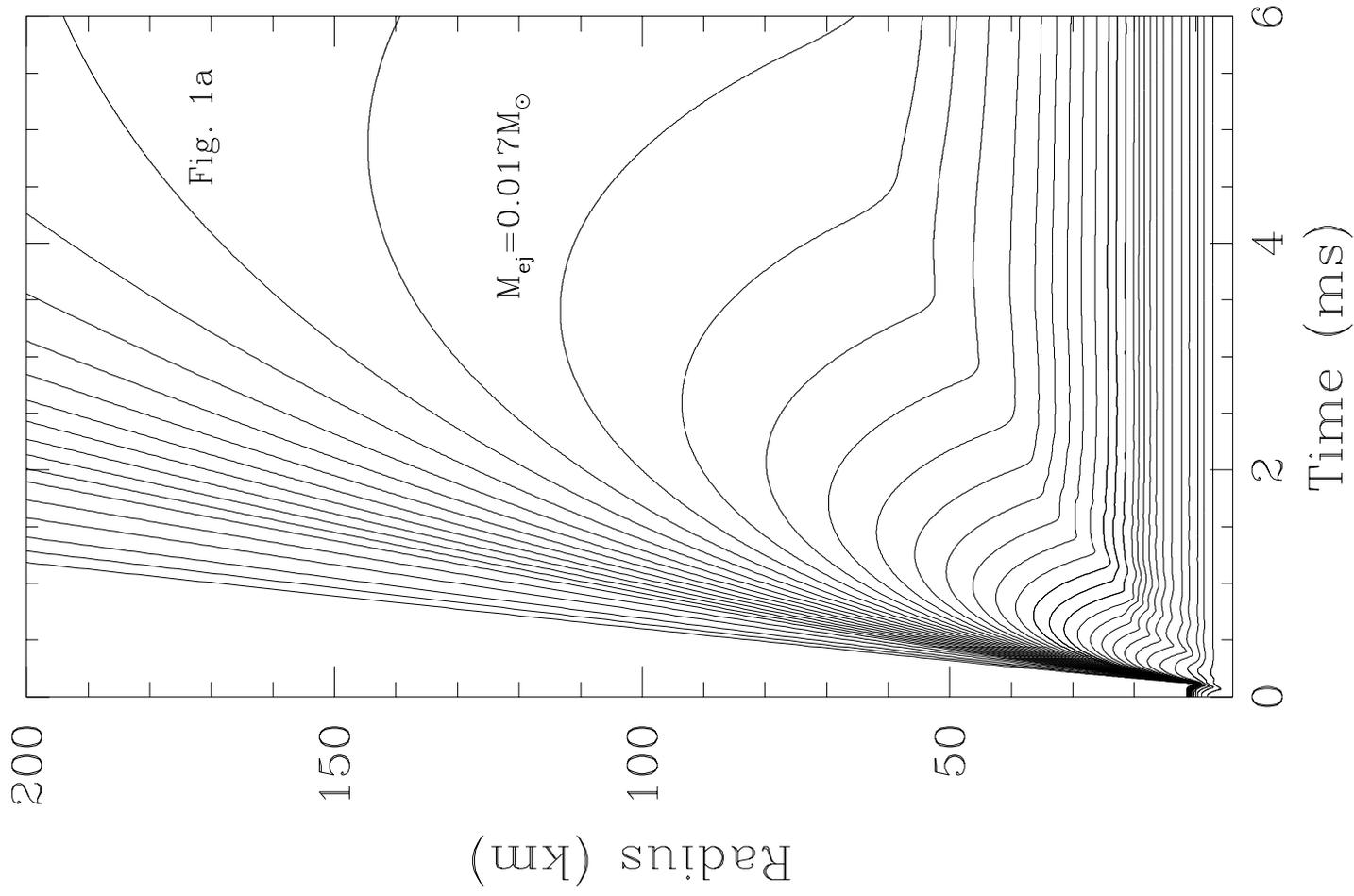
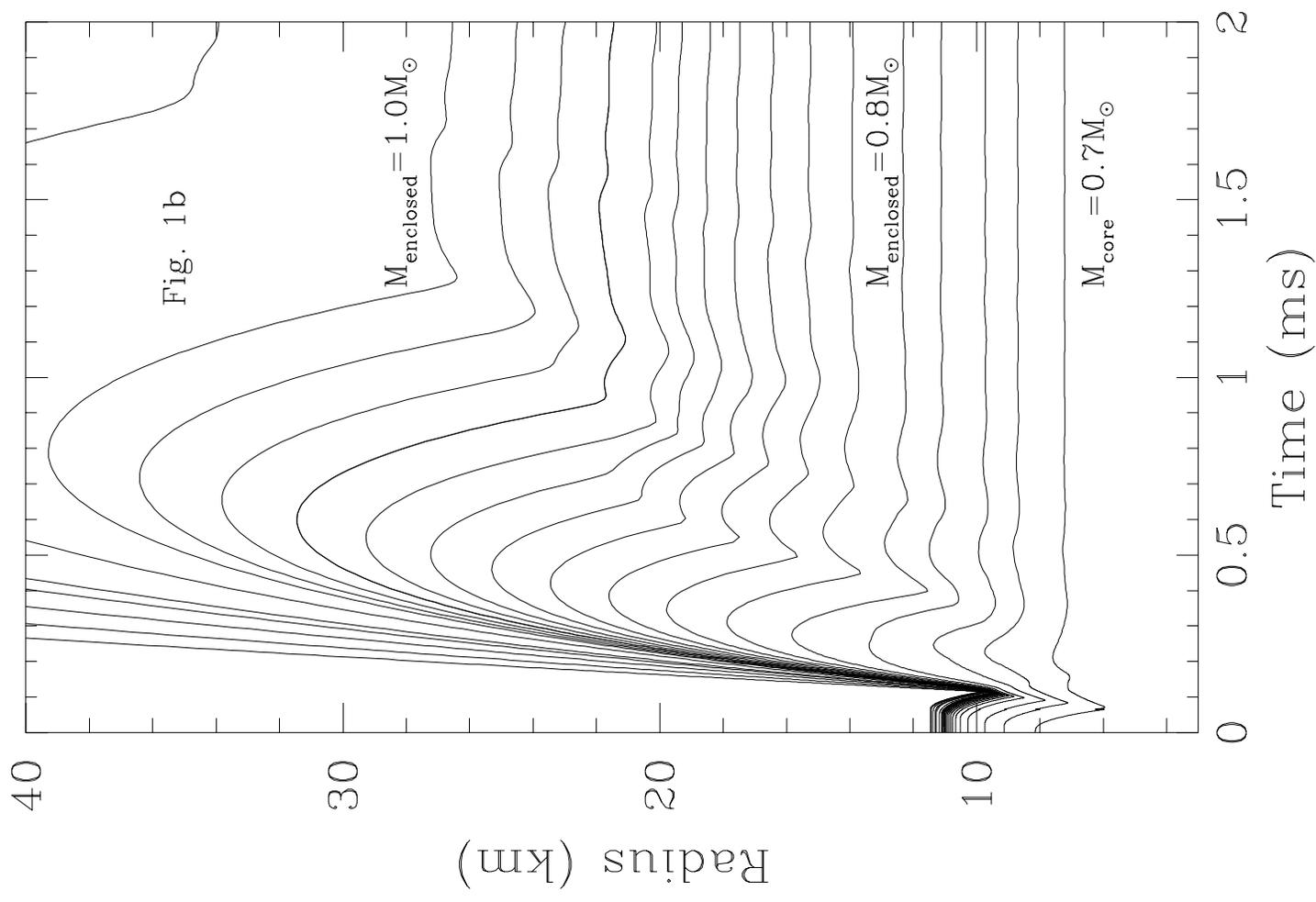
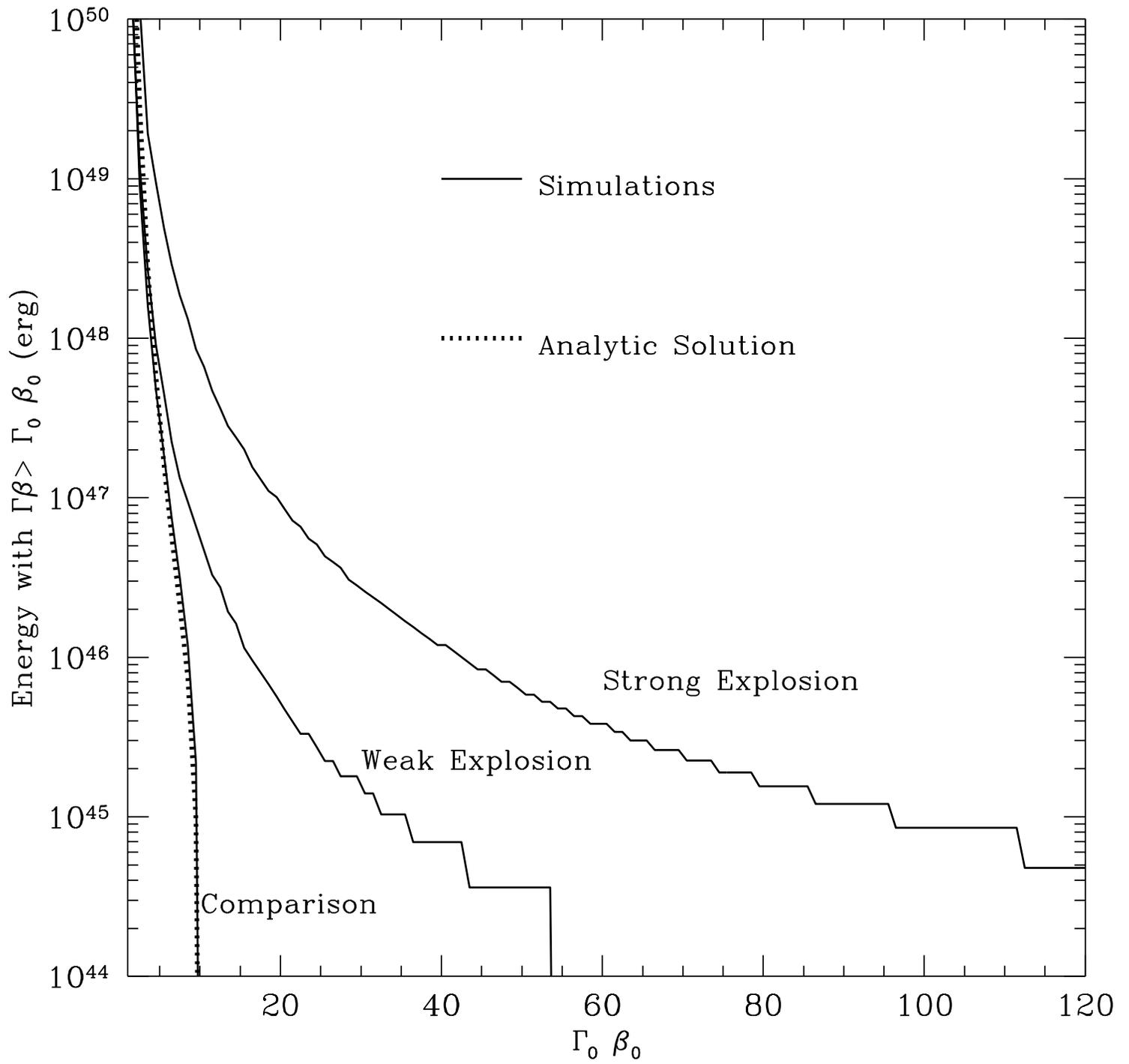
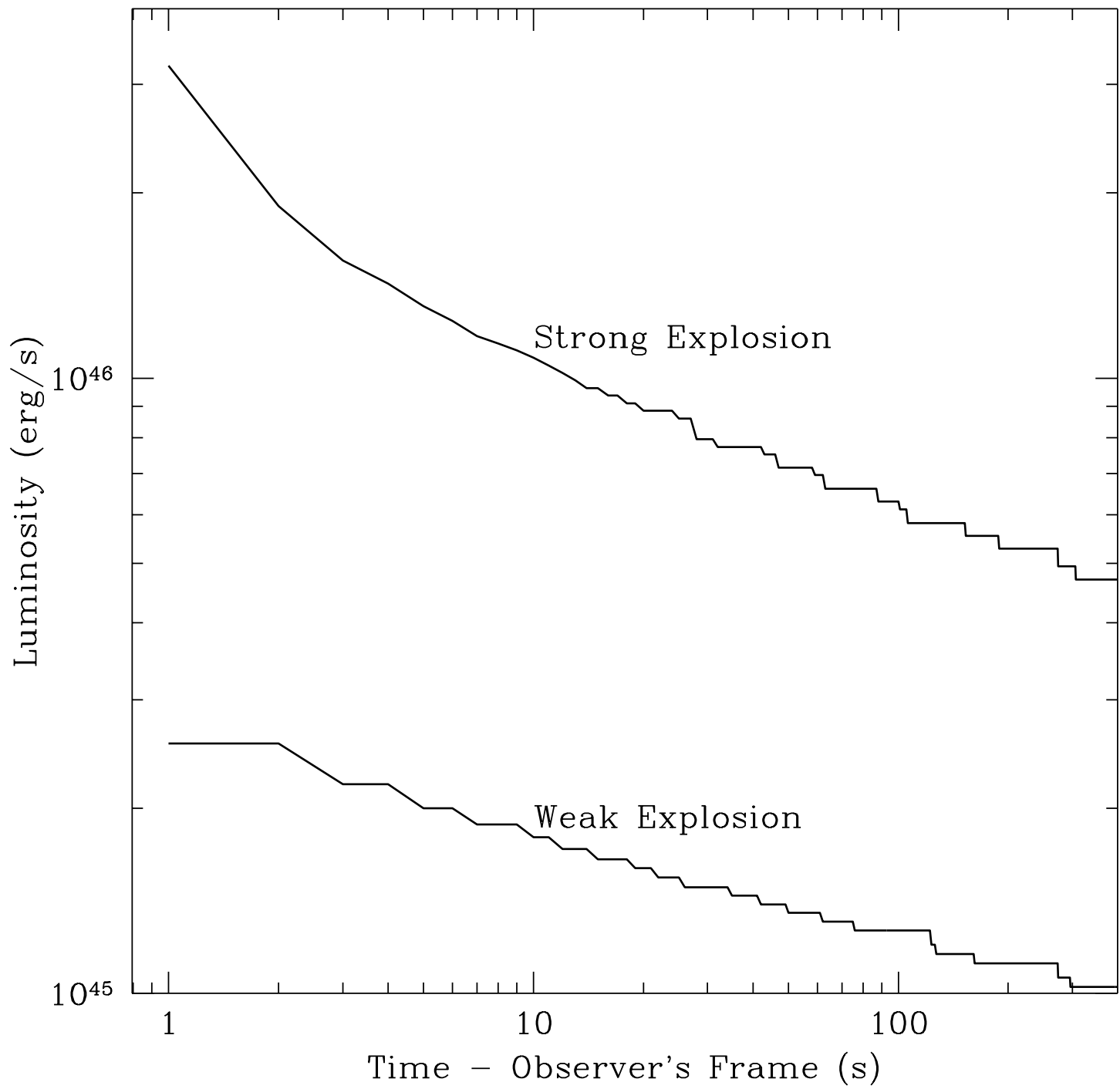
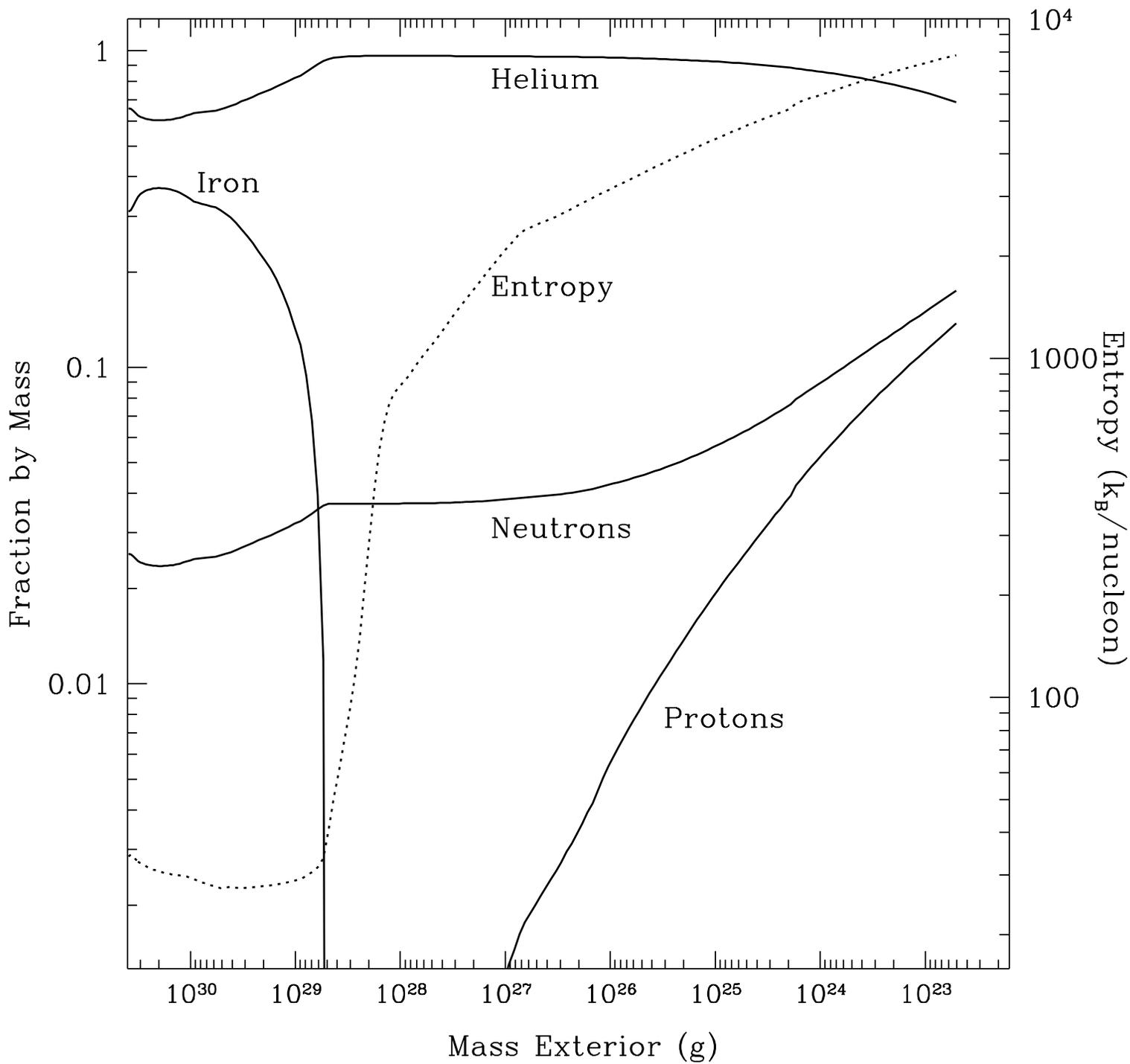
\end{document}